\documentclass[twocolumn,showpacs,amsmath,amssymb,aps]{revtex4}

\usepackage{graphicx}
\usepackage{dcolumn}
\usepackage{bm}

\begin{document}


\title{Depolarization field in thin ferroelectric films with account of semiconductor electrodes\\}

\author{M. D. Glinchuk}
 \email{glin@materials.kiev.ua}
\author{B. Y. Zaulychny}%
 \email{zaulychny@ukr.net}

\affiliation{%
Institute for Problems of Materials Science, National Academy of Sciences of Ukraine\\
}%

\author{V. A. Stephanovich}
\affiliation{
Institute of Mathematics and Informatics, University of Opole, 45-052 Opole, Poland\\
}%

\date{\today}

\begin{abstract}
Within the framework of the phenomenological Ginzburg-Landau theory influence of semiconductor electrodes on the properties of thin ferroelectric films is considered. The contribution of the semiconductor electrodes with different Debye screening length of carriers is included in functional of free energy.\\ 
\indent The influence of highly doped semiconductor electrodes on the depolarization field and the film properties was shown to be great.
\end{abstract}

\maketitle

\indent Influence of electrodes on properties of thin ferroelectric films attracts much attention from scientists and engineers. It is related to substantial influence of electrodes on the field of depolarization, and also with the necessity of proper electrodes types (superconductor, metal, semiconductor) choice which is optimal for applications. The field of depolarization plays essential role in the physics of ferroelectrics, because it tries to destroy spontaneous electric polarization and so ferroelectric phase. It is known that such internal factors as domain structure and free carriers partly diminish the field of depolarization. The external factors, namely electrodes, can substantially decrease the field of depolarization. For example, superconductive electrodes in the bulk ferroelectrics lead to complete compensation of the depolarization field. In thin ferroelectric films due to inhomogeneity of the polarization related to the contribution of surface effects, there is only partial compensation of the depolarization field even for superconductive electrodes \cite{BinderKretchmer}. The account of semiconductor electrodes influence resulted in the necessity of including the contribution of electrodes into free energy, which appeared considerably more difficult for the case of non-superconductive electrodes (see \cite{tilley} and references therein). Because of this up to now the calculations of non-superconductive electrodes influence on the properties of thin ferroelectric films were not carried out.\\
\indent In this work we made these calculations in the model of monodomain ferroelectric material treated as an ideal insulator. This model is realistic enough, as with diminishing of film thickness it becomes a monodomain one \cite{bratkovsky} and in majority of ferroelectrics conductivity is very small (see for example \cite{suchaneck}).\\
\begin{figure}
 \includegraphics{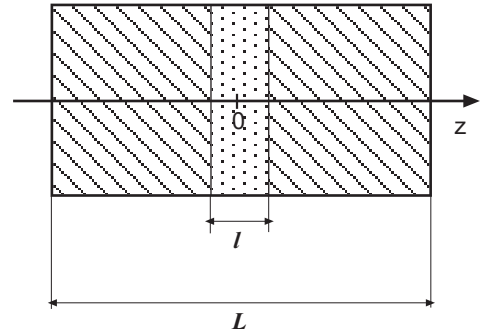} 
 \caption{\label{fig1}Geometry of the film with electrodes. Electrodes (area shaded by slants), film (area shaded by points).}
\end{figure}

\indent We will consider thin ferroelectric film between two highly doped semiconductor electrodes (Fig.~\ref{fig1}). Taking into account that monodomain films selfpolarized in the normal direction to the surface technologically can be produced \cite{pike,bruchaus}, we will examine the film polarized along the axis of z (i.e. $P=P_z\ne 0, P_x=P_y=0$).\\
\indent The equilibrium values of polarization can be obtained within the framework of phenomenological theory from the condition of free energy functional minimum \cite{frydkin}. We will write down free energy as a sum of free energy of film, including the field of depolarization, and electrodes. It is known that the field of depolarization is oppositely directed to spontaneous polarization and so aims to destroy it. In the accepted model of monodomain film without the carriers only charge carriers in electrodes can decrease the field of depolarization. This effect is maximal for superconductive electrodes. For semiconductor electrodes the effect of band bending has to be taken into account \cite{Anselm}, so that the potential of the field in an electrode satisfies the Poisson's equation:
\begin{subequations}
\label{eq:field screening}
 \begin{equation}
  \label{subeq:field screening-law}
  -\frac{{d^2 \varphi }}{{dz^2 }} = \frac{4\pi e}{{\varepsilon_e  }}[n_D-n(z)]
 \end{equation}
where $n_D,~n(z),~e,~\varepsilon_e$ - density of ionized donors, density of free electrons, charge of carrier and background dielectric constant of semiconductor electrode. Due to low density of carriers ($n\approx 10^{16} - 10^{19}$) for it's density Bolzman's law has place: $n(z)=n\cdot exp[{e \varphi}/{kT}]$. Because of semiconductor electrically neutral $n(z)=n_D=n$ in bulk of electrode. Let introduce carrier's potential: $V=-e\varphi$, so than:
 \begin{equation}
  \label{subeq: Carriers potential diff.eq}
  \frac{{d^2 V }}{{dz^2 }} = \frac{4\pi e^2}{{\varepsilon_e  }}n[1-e^{\frac{-V}{kT}}]
 \end{equation}
Introducing dimensionless potential of carrier  
 \begin{equation}  
  \label{subeq: Carriers potential diff.eq substitution}
  \Phi(z)=\frac{V(z)}{kT}
 \end{equation}
and Debye length 
 \begin{equation}
  \label{subeq:field screening-ls-definition}
  l_D ^2  = \frac{\varepsilon_e kT}{4\pi e^2 n}
 \end{equation}
 We have rewritten Eq.(\ref{subeq:field screening-law}) in form
 \begin{equation}
  \label{potential diff.eq}
  \frac{{d^2 \Phi(z) }}{{dz^2 }} = \frac{1}{l_D^2}[1-e^{-\Phi(z)}]
 \end{equation}
 
\end{subequations}

\indent Exact solution of Eq.(\ref{potential diff.eq}) gives exact distribution of electric field inside electrodes, but it is rather complicated, so if we will assume that $\Phi \ll 1$ i.e. $V(z) \ll kT$, thus $exp[-\Phi(z)]\approx 1-\Phi(z)$ after simplification we obtain more simple differential equation:
 \begin{equation}
  \label{simple potential diff.eq}
  \frac{{d^2 \Phi(z) }}{{dz^2 }} = \frac{1}{l_D^2}\Phi(z)
 \end{equation}

\indent It is seen that because of the wide values interval of $n$ in semiconductors, value of $l_D$ can change from units of angstrom to hundreds of angstrom. One can expect that with increase of $l_D$ contribution of electrodes to decrease of the depolarization field will diminish.\\
\indent For quantitative consideration of electrodes influence on film properties we will find the contribution of electrodes to the density of free energy. For this we will search for solution of Eq.(\ref{simple potential diff.eq}) where substitution $E_e=-\frac{d\Phi}{dz}$ has been made, with respect to Maxwell's equation and boundary conditions on electrode's surface:
\begin{subequations}
\label{eq:maxwell}
 \begin{equation}
  D_e=\varepsilon_e E
 \end{equation}
 \begin{equation}
 \label{subeq:maxwell1}
  divD_e= 4 \pi \rho
 \end{equation}
 \begin{equation}
 \label{subeq:maxwell boundary condition1}
  E_e(-\frac{L}{2})=0
 \end{equation}
 \indent Integration of Eq.(\ref{subeq:maxwell1}) leads to new boundary condition:
 \begin{equation}
 \label{subeq:maxwell boundary condition2}
  E_e(-\frac{l}{2})-E_e(-\frac{L}{2})=\frac{4 \pi Q}{\varepsilon_e}
 \end{equation}
 where $Q=\int \limits_{ - \frac{L}{2}}^{-\frac{l}{2}} {\rho dz}=\int \limits_{ \frac{l}{2}}^{\frac{L}{2}} {\rho dz}$ is surface density of charge that accumulated on right-hand surface of left-hand electrode or on left-hand surface of right-hand electrode.
\end{subequations}

\indent Thus, solution of Eq.(\ref{subeq:field screening-law}) for left-hand electrode and right-hand will be:
\begin{subequations}
\label{eq:Ee}
 \begin{equation}
 \label{subeq:lEe}
 ^l E_e \left( z \right) = \frac{{4 \pi Q  \sinh \left( {\frac{{2z + L}}{{2l_D }}} \right)}}{{ \varepsilon _e \sinh \left( {\frac{{L - l}}{{2l_D }}} \right)}}
 \end{equation}
 \begin{equation}
 \label{subeq:rEe}
 ^r E_e \left( z \right) = - \frac{{4 \pi Q  \sinh \left( {\frac{{2z - L}}{{2l_D }}} \right)}}{{ \varepsilon _e \sinh \left( {\frac{{L - l}}{{2l_D }}} \right)}}
 \end{equation}
\end{subequations}

\indent Let's find electric field inside ferroelectric film using Maxwell's equation, condition of $D$ continuity on the interface and the model assumption that $\rho=0$ inside the film:

 \begin{equation}
 \label{Ef eq}
  divD_f=0 ~and~so~ \frac{dD_f}{dz} = 0
 \end{equation}
 With respect to $D_f=E_f+4 \pi P$ the solution of Eq.(\ref{Ef eq}) gives:
 \begin{equation}
 \label{Ef solution}
  E_f=E_0-4 \pi P
 \end{equation}
 The condition of $D$ continuty at the boundary $z=-l/2$ yields:
 \begin{equation}
 \label{E0 to Q}
   E_0=4 \pi Q
 \end{equation}

\indent Now we can summarize behaviour of electric field within the whole system:
\begin{equation}
\label{eq:Ez}
E\left( z \right) = \left\{ {\begin{array}{*{20}c}
   {^l E_e }  \\
   ~\\
   {E_0  - 4 \pi P }  \\
   ~\\
   {^r E_e }  \\
\end{array}\begin{array}{*{20}c}
   ,  \\
   ~\\
   ,  \\
   ~\\
   ,  \\
\end{array}\begin{array}{*{20}c}
   { - \frac{L}{2} \le z \le  - \frac{l}{2}}  \\
   ~\\
   { - \frac{l}{2} \le z \le \frac{l}{2}}  \\
   ~\\
   {\frac{l}{2} \le z \le \frac{L}{2}}  \\
\end{array}} \right.
\end{equation}

\indent In order to find $E_0$ we must account for influence of external voltage $V_0$, namely:
\begin{equation}
\label{eq:ext voltage}
 \int\limits_{- L/2}^{ L/2} {E\left( z \right)dz}  =  - V_0 
\end{equation}

\indent After substitution of Eq.(\ref{eq:Ez}) into Eq.(\ref{eq:ext voltage}) we obtain $E_0$ in the form:
\begin{equation}
\label{eq:E0}
\begin{array}{l}
 E_0 \left( z \right) = \frac{1}{{\left( {2\alpha l_D  + l} \right)}}\left[ {4 \pi \int\limits_{ - l/2}^{l/2} {Pdz}  - V_0 } \right] \\ 
  \\ 
 \alpha  = \frac{1}{{\varepsilon _e }}\frac{{\left( {\cosh \left( {\frac{{L - l}}{{2l_D }}} \right) - 1} \right)}}{{\sinh \left( {\frac{{L - l}}{{2l_D }}} \right)}} \\ 
 \end{array}
\end{equation}

\indent The obtained expression for $E_0$ defines completely electric field in the film (see Eq.(\ref{Ef solution})) and in the electrodes (see Eqs.(\ref{E0 to Q}),(\ref{eq:Ee})).\\
\indent We can express density of free energy of electrodes as
\begin{equation}
 \begin{array}{l}
  F_e=\frac{2}{L-l} \int\limits_{ - L/2}^{ - l/2} {\frac{{D^l E_e }}{{8\pi }}dz}=\frac{{2 \pi Q^2 l_D \beta }}{{(L-l)}}=\frac{{E_0^2 l_D \beta }}{{8 \pi (L-l)}} \\ 
  \\ 
  \beta  = \frac{1}{2}\frac{{\sinh \left( {\frac{{L - l}}{{l_D }}} \right) - \frac{{L - l}}{{l_D }}}}{{\varepsilon _e \sinh ^2 \left( {\frac{{L - l}}{{2l_D }}} \right)}} \\ 
 \end{array}
\end{equation}

\indent The density of free energy  of the film can be written in conventional form for the ferroelectrics with the second order phase transition:
\begin{equation}
\label{Ff}
 \begin{array}{l}
  F_f  = \frac{1}{l}\int\limits_{ - l/2}^{l/2} {dz\left\{ {\frac{1}{2}AP^2  + \frac{1}{4}BP^4  + \frac{1}{2}C\left( {\frac{{dP}}{{dz}}} \right)^2  - } \right.}  \\ 
  \\ 
  - \left. {E(z)P-2 \pi P^2} \right\} + \frac{{C\delta ^{ - 1} }}{{2l}}\left( {P^2 \left( { - \frac{l}{2}} \right) + P^2 \left( {\frac{l}{2}} \right)} \right) \\ 
 \end{array}
\end{equation}

\indent Here $A=A_0(T-T_c)$, $T_c$ and $A_{0}$ is a temperature of ferroelectric transition in the bulk and the inverse constant of Curie-Weiss respectively, $\delta $ is extrapolation length. \\
\indent Finally, after substitution Eqs.(\ref{eq:E0}, \ref{eq:Ez}) into Eq.(\ref{Ff}) we get explicit expression for total density of free energy of the system:
\begin{equation}
 \label{eq:explicit F}
\begin{array}{l}
 F = \frac{{F_e S(L - l) + F_f Sl}}{{SL}} =\\
 ~\\
  = \frac{l}{L}\left[ {\frac{1}{l}\int\limits_{ - l/2}^{l/2} {dz\left\{ {\frac{1}{2}AP^2  + } \right.} } \right.\frac{1}{4}BP^4  + \frac{1}{2}C\left( {\frac{{dP}}{{dz}}} \right)^2  + ~~~~~~~~~~~~ \\ 
  \\ 
  + \frac{{V_0 P}}{l}\left( {\frac{l}{{\left( {2\alpha l_D  + l} \right)}} - \frac{{l_D \beta l}}{{\left( {2\alpha l_D  + l} \right)^2 }}} \right) - \frac{4 \pi \bar P P l}{{\left( {2\alpha l_D  + l} \right)}} +  \\ 
  \\ 
 \left.  + 2 \pi P^2 \right\} \left. { + \frac{{C\delta ^{ - 1} }}{{2l}}\left( {P^2 \left( { - \frac{l}{2}} \right) + P^2 \left( {\frac{l}{2}} \right)} \right)} \right] +  \\ 
  \\ 
  + \frac{{2 \pi \left( {\bar P} \right)^2 l_D \beta l^2 }}{{ L\left( {2\alpha l_D  + l} \right)^2 }} + \frac{{V_0 ^2 l_D \beta }}{{8 \pi L\left( {2\alpha l_D  + l} \right)^2 }} \\ 
 \end{array}
\end{equation}

\indent Variation of functional (\ref{eq:explicit F}) leads to the Euler-Lagrange equation for polarization and to boundary conditions of the following form:
\begin{subequations}
 \begin{equation}
  \label{eq:Euler equation}
  \begin{array}{l}
 AP + BP^3  - C\left( {\frac{{d^2 P}}{{dz^2 }}} \right) = E_{ext}  + E_d  ~~~~~~~~~~~~~~~~~~~~~\\ 
  \\ 
 E_{ext}  =  - \frac{{V_0 }}{l}a{\rm~~~~~~~~~~~}E_d  =  - 4 \pi \left( {P - a\bar P} \right) \\ 
  \\ 
 a = l \left( {\frac{1}{{\left( {2\alpha l_D  + l} \right)}} - \frac{{l_D \beta}}{{\left( {2\alpha l_D  + l} \right)^2 }}} \right) \\ 
  \end{array}
 \end{equation}
 \begin{equation}
  \left. {\frac{{dP}}{{dz}}} \right|_{z =  \pm \frac{l}{2}}  =  \mp {\textstyle{{P\left( { \pm \frac{l}{2}} \right)} \over \delta }}
 \end{equation}
\end{subequations}

\indent The expression for $E_d$ is the same as for metallic electrodes \cite{BinderKretchmer}, \cite{glin1}, \cite{glin2},  \cite{Zaulychny} where $a\approx 1$. However in general case $a<1$, and for semiconductor electrodes $a$ significantly less than 1 so that depolarization field increase.\\
\indent It is expected that semiconductor electrodes which have large enough $l_D$ value and small $\varepsilon_e$ will lead to increase depolarization field and so to the decrease of spontaneous polarization and to the increase of critical thickness of phase transition from ferroelectric to paraelectric phase. Preliminary calculations had shown that at some critical value of $l_D$ it can be semiconductor substrate induced ferroelectric-paraelectric phase transition, and that all the film properties are dependent on parameter $a$ and so on the characteristics of electrodes. This opens the way of proper electrodes types choice (superconductive, metallic or semiconductor) which is optimal for applications.\\
\begin{acknowledgments}
We are grateful to Prof. D. R. Tilley for drawing our attention to importance of arbitrary metallic electrodes contribution calculations and fruitful discussion of the results.
\end{acknowledgments} 

\bibliography{article}
\end{document}